\documentclass[aps,prb,twocolumn,english]{revtex4-2} 

\usepackage{amssymb}
\usepackage{amsmath} 
\usepackage{mathtools}
\usepackage{epsfig}
\usepackage{color}
\usepackage[normalem]{ulem} 

\usepackage[unicode=true,pdfusetitle,
 bookmarks=true,bookmarksnumbered=false,bookmarksopen=false,
breaklinks=false,pdfborder={0 0 0},pdfborderstyle={},backref=false,colorlinks=true]
 {hyperref}

\newcommand{\be}{\begin{equation}}
\newcommand{\ee}{\end{equation}}
\newcommand{\bea}{\begin{eqnarray}}
\newcommand{\eea}{\end{eqnarray}}

\newcommand{\p}{\partial}
\newcommand{\s}{\sigma}

\newcommand{\la}{\langle}
\newcommand{\ra}{\rangle}
\newcommand{\rd}{\mbox{d}}
\newcommand{\ri}{\mbox{i}}
\newcommand{\re}{\mbox{e}}

\begin{document}

\title{Two Channel  Multi impurity Kondo  model: RKKY induced criticality}
\author{ A. M. Tsvelik}
\affiliation{Division of Condensed Matter Physics and Materials Science, Brookhaven National Laboratory, Upton, NY 11973-5000, USA}
\author{G. Kotliar}
\affiliation{Department of Physics and Astronomy, Rutgers University, 136 Frelinghuysen Rd., Piscataway, NJ 08854-8019, USA}

\date{\today}

\begin{abstract}
We show  that the  Ruderman-Kittel-Kasua-Yoisida  interaction  between overscreened spins in  two channel Kondo impurity systems  is a relevant perturbation  when the number of impurities N  is greater than 3  driving the system to a new quantum critical point   with anomalous dimensions  $\frac{1}{ (N+1)} $  for the spin operator 
and  the  Sommerfeld coefficient of the specific heat  scales  as $\gamma \sim T^{- \frac{3}{N+1}}$. 
The critical point universal properties are  relevant to many strong correlation problems,  such as  impurity placed in a Majorana metal  and   the multichannel Kondo lattice model of heavy fermion materials. We discuss relevance of our results for cluster DMFT  studies of quantum criticality. 
\end{abstract}

\maketitle

\textbf{Introduction. }
The study of quantum impurity models has played a crucial role in the development of modern theories of strongly correlated electron systems\cite{Hewson1993}.  Beginning with Anderson's single impurity model and Wilson's analysis of the renormalization group of the Kondo problem, these models have provided profound insights into the emergence of nontrivial many-body phenomena such as spin screening, quantum entanglement, and the breakdown of Fermi liquid theory. Theoretical advances in impurity models have not only led to fundamental concepts such as the Kondo effect, heavy-fermion behavior, and quantum criticality but have also been instrumental in approaching  the many body problem on the lattice, as for example in  dynamical mean-field theory (DMFT), which treats lattice problems by mapping them onto effective  impurity models.


Quantum impurity models are also important from the perspective of 
unconventional quantum critical points that are  not well described by the Hertz-Millis-Moriya theory of quantum criticality. These zero temperature  critical points are  characterized by charge or spin susceptibilities that exhibit weak momentum dependence, while their frequency dependence follows a logarithmic form over a broad frequency range,   and a  specific heat behaving as $  C(T)\sim T\ln T $ reminiscent of the two-channel Kondo impurity model.   Salient examples include the marginal Fermi liquid theory describing optimally doped cuprates \cite{PhysRevLett.63.1996} and the quantum critical point of heavy fermion systems for which the term local quantum criticality \cite{Si2001}  is  widely used.   Interestingly, while the two-channel Kondo impurity model captures the logarithmic scaling of various observables over a broad  range of energies, a more singular behavior emerges at much lower temperatures or frequencies in both heavy fermions  \cite{OESCHLER20081254} and cuprates while still being  conformally invariant as expected from a quantum impurity model
\cite{CFTguo2024conformallyinvariantchargefluctuations}.

The study of models with multiple  impurities,  brings  new features  arising from the RKKY interaction which  have motivated 
experiments in mesoscopic systems as well as molecules absorbed on surfaces \cite{Experiment1,Experiment2,Spinelli}. 
\cite{Goldhaber}. See Ref. \cite{Eickhoff2021} for a recent context and a proper identification of the number of channels. 

 The standard paradigm suggests that the  competition between the RKKY interaction and the Kondo screening  results in a quantum critical point  when these two scales are comparable.  This scenario has been explored 
by Jones and Varma \cite{Varma}, whose work  was later generalized by Georges and Sengupta \cite{georges, AnKondo1} to the  two channel situation. The three impurity  Kondo model, displays more a complex quantum critical point,  and was studied in Ref. \cite{PhysRevLett.95.257204}.  
 In this letter we depart from this scenario and show  that in the two channel Kondo model the RKKY interaction itself can generate new types of quantum criticality  even when it is much smaller than the single impurity two channel Kondo scale.
We solve exactly  a model of  the  spin sector of the multichannel N impurity model,  and show that it  exhibits qualitatively new   critical features   when    the number of impurities $ N\geq 3 $.  We study a  QIM (quantum impurity model), in the same universality class of the topological Kondo model      studied in  refs. \cite{BeriCooper, AlEg, AlTsv, AlTsv2,Jukka}. It  exhibits  a  non trivial fixed point, arising directly from the presence of the RKKY interaction.  The model exhibits two channel Kondo behavior at intermediate energies and crosses over to a more singular behavior at lower energies making it a possible candidate for the local quantum critical point governing the heavy fermions and the cuprates.  This QIM has  many  other  interesting   potential  applications and we  illustrate this with some implications to    the physics of 
magnetic impurities embedded in a Majorana metal and  to   
 heavy fermion materials, which host two channel  Kondo  lattice physics such as
  UBe$_{13}$\cite{cox1994resistivitytwochannelkondomodel} and  PrV$_{2}$Al$_{20}$ \cite{PhysRevRes6},\cite{Nakatsuji}.

\textbf{Construction of the effective theory.}
We start with the theory of a  single spin coupled to  two channels of conduction electrons in three  dimensions  via an isotropic spin interaction. 
 It is well established  that when the channels is coupled symmetrically a part of the bulk degrees of freedom decouple from the impurity such that the resulting effective Hamiltonian can be recast in terms of the three species of Majorana fermions.  The two channel Kondo effect becomes quantum critical. The stability of the critical point may be achieved for magnetic ions with non-Kramers ground state such as Pr ones \cite{PhysRevRes6}. The Kondo temperature $T_K$ is exponential in $1/J$  and the effective  Hamiltonian  $E << T_K$ 
 is given by \cite{Ioffe} 
\bea
H_{eff} = \sum_k \epsilon(k)\Big(\vec\chi(-k)\vec\chi(k)\Big) +  T_K^{-1/2} \xi(\chi^1\chi^2\chi^3)_0 \label{effH}
\eea

At low energy, and to  the leading order  the low energy excitations are described by three non interacting Majorana fields $\chi^a$. 
 $\xi$ is  a  drone variable,  a   local Majorana zero mode, which enables the representation of the impurity spin operator  at  energies $<< T_K$ as:
\bea
\vec S\rightarrow \ri (\Lambda/T_K)^{1/2}\xi \vec\chi(x=0) +... \label{S1}
\eea
$\chi^a$ denotes the band Majorana fermion located at the impurity site, 
$\Lambda $    the UV cut-off is  of the order of the Fermi energy and the dots stand for less relevant operators.

 As far as the impurity is concerned one can replace the 3D fermions with 1D chiral ones,  then the  effective action corresponding to the Hamiltonian  (\ref{effH}) is 
\bea
&& A_{2ch} = \int \rd\tau\int_{-\infty}^{\infty}\rd x \frac{1}{2}\Big(\chi^a\p_{\tau}\chi^a - \ri\chi^a\p_x\chi^a\Big) + \frac{1}{2}\xi\p_{\tau}\xi + \nonumber\\
&& T_K^{-1/2}\xi(\chi^1\chi^2\chi^3)_0 \label{action}
\eea
and  $x$  is   a  coordinate in one dimension.

\begin{figure}
\centering
\includegraphics[width=0.6\linewidth]{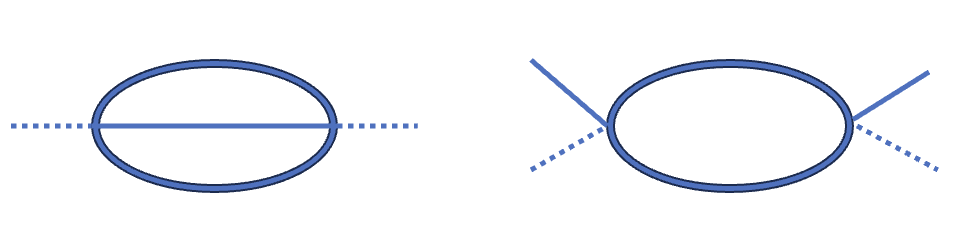}
\caption{\label{fig:fusion} The diagrams involved in the generation of fusion (\ref{interaction}). The dotted lines correspond to drone fermions $\xi$, the solid lines are Green's functions of the itinerant Majoranas.}
\end{figure}

 However, when considering many impurities one should not forget that the fermions screening different impurities belong to the same D dimensional bulk so that their correlation functions do not vanish.  We will take this fact into account by integrating 
 out part of the fermionic fields not located at the impurity sites. To this end we use Hamiltonian (\ref{effH}) and perform fusion of the four-fermion terms located at different sites: 
 \begin{eqnarray}
 && {\cal O}(r,\tau) \coloneqq  T_K^{-1/2}[\xi\chi^1\chi^2\chi^3](r,\tau),   \nonumber\\
 && \int \rd\Delta\tau{\cal O}(\tau, r_1) {\cal O}(\tau +\Delta\tau, r_2) = \nonumber\\
 && \Big[  H(R_{12})\xi_1\p_{\tau}\xi_2 + \bar I(R_{12})\xi_1\xi_2 ({\vec \chi}_1{\vec\chi}_2) +... \Big], \label{interaction}
 \end{eqnarray}
 where the dots stand for less relevant operators.  The fusion is determined by the Feynman diagrams depicted on Fig. \ref{fig:fusion}.  
The fusion parameters are 
\bea
 &&H_{ij} = (\Lambda/T_K)\int\rd\tau  \tau G^3(\tau, r)\nonumber\\
 && \bar I_{ij}= (\Lambda/T_K)\int \rd\tau  G^2(\tau,r)
 \label{RKKY2}
 \eea
 The exact forms of  $H(R)$ and the RKKY interaction $I(R)$ depend on the dimensionality and the band structure. Being a linear combination of particle and hole the Majorana Green's functions is 
 \bea
 G(\omega_n, k) = \frac{1}{2}\Big[\frac{1}{\ri\omega_n - \epsilon(k)} + \frac{1}{\ri\omega_n + \epsilon(k)}\Big].
 \eea
 For the spherical 3D Fermi surface the result is  
 \bea
 G(\tau,\vec r) =  \frac{\rho(\epsilon_F)\tau}{\tau^2 + (r/v_F)^2}\frac{\sin(k_Fr)}{k_Fr}.
 \eea
where $k_F$ is the Fermi wave vector, $\rho(\epsilon_F)$  the density of states at the Fermi level and $r_{ij}$ is the distance between the impurity sites.
We see that since $G(\tau,R) = -G(-\tau,R)$ the tunneling of the drone fermions vanishes at $\omega =0$ resulting in the time derivative in (\ref{interaction}).

 We see that the fusion of the operators which are irrelevant for the single impurity fixed point  generates a strictly marginal and a marginally relevant ( $I >0$) operators.  However, one has to remember that the fusion procedure is self consistent only in the dilute limit when  the resulting interactions affect only the low energy sector of the model determined by the upper cut-off $T_K$. Hence we must have 
 \be
 H << T_K, ~~\rho(\epsilon_F)I << T_K.\label{conditions}
 \ee
 This is a strong self-consistency requirement which will be discussed later. Under these assumptions we will neglect the $H$-term altogether. The resulting low energy ($E< T_K$) multi-impurity action is a Kondo model where one can effectively treat the bulk as a set of independent  one-dimensional wires: 

\bea
 A = \sum_n^N A_{2ch}\Big[\vec\chi_n, \xi_n\Big] + \sum_{n,m} \int \rd\tau \bar I(R_{nm})\xi_n\xi_m ({\vec \chi}_n{\vec\chi}_m) \label{multi}
 \eea
 with $A_{2ch}$ is given by (\ref{action}). The ultraviolet cut-off is $v/r$, where $r$ is an average distance between the impurities. 
 The resulting model  is a multichannel version of the model for the topological Kondo effect studied in \cite{BeriCooper, AlEg, AlTsv, AlTsv2}. 
 
 The multichannel version (\ref{multi}) was recently studied  in \cite{Jukka}, 
  where it was established that it  is critical. Therefore  contrary to the conventional wisdom, the criticality is generated rather than destroyed by the RKKY interaction, at least  provided the impurity concentration is sufficiently small so that the conditions (\ref{conditions}) are fulfilled.


We now discuss  the physics resulting from the effective action (\ref{multi})  for different number of impurities N.  We also allow for the site dependence  anisotropy in the RKKY interactions  $\bar I_{nm}$.

\textbf{ N=2 impurity case. }
For N=2, 
 we considered the simplest model which takes into account only a direct interaction between the spins: 
\bea
H = H_{2ch1} + H_{2ch2} + \bar I^a  S^a_1S^a_2. \label{A}
  \eea
  a special case of the two-impurity models considered in \cite{ingersent,georges} which  included many couplings related to scattering between the impurities.  

  
Already this model is sufficient to describe  a crossover from the 2-channel QCP to a QCP with more  a more singular magnetic  response. 
  
  At $I^a << T_K$  for the spins we can adopt their expressions at the 2-channel QCP (\ref{S1}).  
  Then the interacting term (\ref{multi}) becomes  
  \bea
   \bar I^a\xi_1\xi_2 \chi^a_1\chi^a_2 = I^a(\Lambda/T_K) (d^+d -1/2):\psi_a^+\psi_a: \label{marginal}
  \eea
  where $\psi = \chi_1 +\ri\chi_2$.  It is strictly marginal and hence continuously changes  the scaling dimensions of the 2-channel Kondo model operators. 


\textbf{Three impurities. } 
We now turn to the three impurities case.  We work at energies much smaller than the single impurity Kondo temperature $T_K$ such that the UV cut-off for the theory is $T_K$. 
  In this case (\ref{multi}), extended to the spin anisotropic case, is equivalent to the Hamiltonian:
 \bea
 && H = T_K^{-1}\sum_m \xi_m(\chi^1_{m}\chi^2_m\chi_m^3)_{x=0} + \label{3imp}\\
 && \sum_{n=1}^3\sum_{a=1}^3 \int \rd x (\ri/2)\chi_n^a\p_x\chi_n^a + \ri \bar I^n \Big[\epsilon_{nmk}\s^n\Big(\vec\chi_m\vec\chi_k\Big)_{x=0}\Big].\nonumber 
 \eea
 where $\s^n =\frac{ \ri }{2}\epsilon_{nmk}\xi_m\xi_k$ and $\bar I$ contains factor $(\Lambda/T_K)$ as in (\ref{RKKY2}). 


   The first term in the Hamiltonian (\ref{3imp}) is irrelevant, but it prevents the model from being exactly solvable in the entire range of temperatures from $T_K$ to zero. The exact Bethe ansatz solution is possible for $|\bar I_1| = |\bar I_2| < \bar I_3$ only asymptotically at $T<< T_K$. The solution is similar to the one for spin $S$ XXZ Heisenberg Hamiltonian \cite{AnKondo2} and has been discussed in \cite{AnKondo3,AnKondo4}.  We discus the results below, for more details on the 
  the Bethe ansatz treatment
  see Sec.II of the Supplemental Material.

 
 In contrast with the $N=2$ case the  RKKY  interaction   is not marginal. It follows from (\ref{RKKY2}) that the sign of the coupling is always positive and hence  the interaction in (\ref{3imp}) is always marginally  relevant. The coupling constants flow under RG as is it customary for non-Abelian theories. Its growth  generates its own crossover scale $T_{RKKY}$.  In the isotropic case we have 
 $
 T_{RKKY} \sim T_K \exp[- 1/2\rho(\epsilon_F)\bar I] 
$.
A trajectory goes from the single impurity two channel QCP SU$_2$(2) at high energies   to the multi-impurity   6-channel QCP at low energies since 
 $O_3(3) \equiv SU_6(2)$, it is the 6-channel Kondo. The thermodynamics can be extracted from \cite{TsvMulti} (see also Supplement).  
From the Bethe anzats solution the  impurity contribution to the ground state entropy is given by 
 $
 S(0) = \ln \sqrt{2 +\sqrt 2}.
 $
The leading  irrelevant operator which determines the temperature dependence of the impurity contribution to the specific heat,  is the first Kac-Moody descendant of the primary field in the adjoint representation $
 J_{-1}\Phi_{adj}$. Its  conformal dimension  at the critical point is $h = j(j+1)/(k+2)$ which for $j=1, k=6$ yields  1/4. 
 The impurity free energy scales as $F_{imp} \sim T^{2h_1+1}$, so the specific heat is  
\bea
C_{imp} \sim (T/T_{RKKY})^{1/2}. \label{irrel}
\eea 
All these results are consistent with the Bethe anzats solution.

\textbf{Renormalization Group Flow Conformal Field Theory and generalization to arbitrary N.}

 As we have stated above, model (\ref{multi}) is a multichannel version of the model describing the  topological Kondo effect studied in \cite{BeriCooper, AlEg, AlTsv, AlTsv2} and was recently studied in \cite{Jukka}.   
 $J_{ij} = \ri  (\vec\chi_i\vec\chi_j)$ are $O_3(N)$ Kac-Moody currents ($N$ is the number of impurities) and $s_{ij} = \ri\xi_i\xi_j$ are generators of the $o(N)$ algebra. Hence we have $n=3$ channel Kondo problem with orthogonal symmetry. 
 
 The bulk Hamiltonian consists of $3N$ Majorana modes with symmetry $O_1(3N)$. Meanwhile the interaction involves $O_3(N)$ Kac-Moody currents which means that only a part of the bulk is involved in the impurity screening. 
We therefore conclude that  the effective Hamiltonian  describing a cluster of $N$ impurities with approximately equal RKKY couplings.
 Channel The CFTtwo analysis can be generalized to an arbitrary number of  Majorana species $n$. Consider the conformal embedding: 
\bea
O_1(nN) = O_n(N)\times O_N(n).
\eea
The $O_n(N)$ sector  absorbs the  impurities and  becomes quantum critical  while the $O_N(n)$ sector of the bulk remains idle. The resulting critical theory is the $O_n(N)$ Wess-Zumino-Novikov-Witten (WZNW) model subject to a conformally invariant boundary condition. Its basic properties are described in \cite{kimura}. The conformal dimensions of the primary fields of the $O_n(N)$ WZNW model are 
$
h_i = \frac{C_i}{n+N-2},
$
where $C_i$ are $O(N)$ group Casimirs in the corresponding representations. 

The most important operators is the operator from the adjoint representation which first Kac-Moody descendant determines a flow towards the ground state. Its conformal dimension is 
$
h_{adj} = \frac{N-2}{n+N-2}.
$
The corresponding contributions to the impurity free energy are 
\bea
F_{imp}^{irrel} \sim -TS(0) - aT(T/T_{RKKY})^{h_{adj}}, 
\eea
where $S(0)$ is the ground state entropy and $a$ is a numerical constant.

  Another important operator is the impurity spin (\ref{S1}). It is a composite operator which includes both sectors. The Majorana fermion $\chi_n^a$ has conformal dimension 1/2 which is a sum of conformal dimensions of the $O_n(N)$ and $O_N(n)$ WZNW models:
$
1/2 = \frac{(N-1)/2}{N+n-2} + \frac{(n-1)/2}{N+n-2}.
$
It is reasonable to assume that the $O_3(N)$ part becomes a singlet with the conformal dimension zero. Then the  remaining part has conformal dimension 
$
h_S = \frac{1}{N+1}.
$
The correlation function 
\bea
\la S^a_n(\tau)S^a_m(0)\ra = F_{nm}\Big|\sin\Big(\pi T\tau\Big)\Big|^{-2/(N+1)}
\eea
is nonlocal in space.

 Our results have important   implications  to many systems of current interest and we illustrate this  with a few  examples:  the Majorana spin liquid and the two channel Kondo lattice.
  

\textbf{Magnetic impurities in a Majorana  spin liquid with a Fermi surface}
  The above results are directly applicable to a  Majorana metal with magnetic impurities. Such metal  is a  spin liquid   where elementary excitations are neutral  Majorana  fermions with a Fermi surface and Z$_2$ gauge fluxes. This situation is realized for certain 3D latticees in an exactly soluble  generalization of the famous Kitaev model of Z$_2$ spin liquid \cite{kitaev} - the Yao-Lee (YL) model \cite{hermanns_physics_2018,obrien_classification_2016,CPT1}.  We provide a more detailed description  in the Supplementary Material. 
 The 3D YL model has a phase transition below which  the static Z$_2$ gauge field can be ignored and   the low energies  excitations are 
 three species of noninteracting Majorana fermions.  The YL model does not have a charge sector. Therefore  a coupling  a cluster of impurity  $S=1/2$  spins to the Yao Lee model with a Fermi surface   provides a straightforward realization of the physics described in this paper.

 The single impurity problem describing spin $S=1/2$ coupled to YL spin liquid with a Fermi surface is equivalent to the spin sector of the two-channel Kondo model. The effective action for the multi-impurity case is given by (\ref{multi}). 

There is one detail worth mentioning. The YL spin field $\s$ is a bilinear in the Majoranas: $\vec\s = \frac{\ri}{2}[\vec\chi\times\vec\chi]$, unlike the situation  in the  conventional Kondo model.
Hence it makes sense  to calculate the self energy of the Majoranas. In the leading order in the impurity concentration (that is at temperatures much larger than $T_{RKKY}$) it is determined by the quartic term in (\ref{effH}): 
\bea
\Sigma \sim n_i \rho(\epsilon_F)T_K^2\int \rd\tau \frac{\sin(\omega\tau)}{\tau^2} \sim n_iT_K^2\rho(\epsilon_F) \omega\ln\omega.
\eea
where $n_i$ is the impurity concentration and $\rho(\epsilon_F)$ is the density of states at the Fermi surface. Recall that for the electrons it would be square root frequency dependence \cite{AffleckLudwig93}. It enters in the expression for the thermal conductivity.  Hence in the limit of small impurity concentration the YL model with spin impurities behaves as  marginal FL with linear T dependence in the thermal resistivity. The low temperature behavior for the multi-impurity case is described above.

\textbf{ Two channel Kondo Lattice Model. }
The  Hamiltonian of the  two channel Kondo lattice,  an archetypical  model in the theory of correlated electron systems is given by 
\begin{eqnarray}
&& H = \sum_{k, m, \sigma} \epsilon_k \, c_{km\sigma}^\dagger c_{km\sigma} + J \sum_{i, m} \mathbf{S}_i \cdot \mathbf{s}_{im}, \nonumber\\
&& \mathbf{s}_{im} = \sum_{\sigma, \sigma'} c_{im\sigma}^\dagger \boldsymbol{\sigma}_{\sigma\sigma'} c_{im\sigma'},
\end{eqnarray}
with  $\epsilon_k$  is the conduction electron dispersion $\sigma$ and $m$ spin and channel indices, running over two values, the index i runs over  the sites of the lattice and $\mathbf{S}_i$  the spin at lattice site i. 
It was proposed as relevant to  the  heavy fermion materials  such as UBe$_{13}$,  which exhibits a large resistivity down to its superconducting transition temperature,  PrRh$_2$Zn$_{20}$ which exhibits quadrupolar order\cite{PhysRevB.87.205106} and PrV$_2$Al$_{20}$ \cite{PhysRevRes6}.  
Not much is known about this model, even in limiting cases. Within  single site DMFT which is exact in infinite dimensions, its normal phase    realizes a non Fermi liquid incoherent metal and  phases with broken symmetry such as  channel  ordering\cite{Motome_PhysRevB.102.155126, Hoshino_doi:10.7566/JPSJ.82.044707,PhysRevRes6}. Slave particle studies  reveal even more exotic properties \cite{Komijani_PhysRevLett.129.077202,PhysRevRes6}. In the paramagnetic phase, cluster and single site  DMFT studies of this model do not exhibit Fermi liquid behavior.   

Our  fixed point is  a candidate  to describe this non Fermi liquid behavior.   DMFT studies this model  in the paramagnetic phase. The approach  maps the lattice problem into a cluster of N  impurities  spins embedded  in a two channel  bath obeying  a    DMFT self consistency condition, as summarized in the Appendix.  In cluster DMFT the impurity self energy,      
$\Sigma_{imp} $ is viewed as a functional of the hybridization function,  and is obtained solving the impurity model. The correlation functions can be computed from the impurity model. 
Our discussion  heavily relies  on the assumption that the self energy of the  type of impurity model  does not diverge at low frequency.  Then as shown in the Appendix,  the self consistent   hybridization function   is regular (i.e. it has a finite density of states)  at low energies in a self consistent way. 
Having a regular  hybridization function,  the results of this paper apply and we expect  a non Fermi  liquid phase, 
starting with  a regime of decoupled impurities at high temperatures, and  flowing to the new fixed point identified in this paper throughout the paramagnetic phase. 

{\bf Conclusions}.

\begin{figure}
\centering
\includegraphics[width=0.6\linewidth]{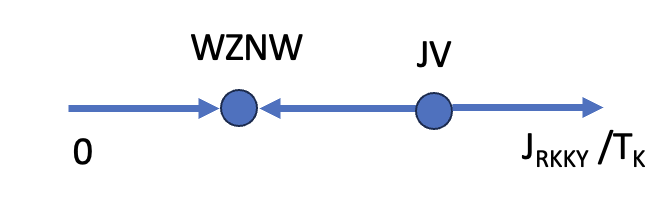}
\caption{\label{fig:RG} The suggested picture of the renormalization group flow for the two channel Kondo model with multiple impurities. There are two critical points: the unstable Jones-Varma one and a new stable Wess-Zumino-Novikov-Witten critical point described in the paper. The phase on the right from the JV-point is magnetic. }
\end{figure}
  We studied  an RKKY driven quantum critical point of a QIM  in the multi-impurity case using a combination of  conformal field theory and Bethe ansatz methods. Our central result is summarized on Fig.\ref{fig:RG}. We suggest that in the diluted regime  when the  RKKY interactions are weak,  the system flows to  a new non Fermi liquid fixed point which controls the physical properties of the system  at low energies.  This critical point is relevant to many problems, such as impurities in thermal metals and multichannel Kondo lattices. 

   The scaling dimension of the spin  operator at the low energy QCP  is $1/(N+1)$ making its correlation function more singular as $N$ increases.  At the same time  the specific heat  singularity  and the impurity entropy per spin  are reduced with the increase of $N$.  This is consistent with the interpretation that  the criticality is connected to freezing of the spin degrees of freedom  due to  the  ordering or  to the formation of  some spin liquid state.  In the context of cluster  DMFT,  the effective number of impurities is determined dynamically, with the effective range determined by the temperature starting from single  impurity behavior at high temperatures\cite{Darko_PhysRevB.84.115105}.

 It would be interesting to relate the  mechanism for criticality  introduced in this paper,  with the observations of criticality found in recent  cluster DMFT studies of the Hubbard model at optimal doping  \cite{Haule_PhysRevB.76.104509,Sordi-PhysRevB.84.075161}, \cite{Gleis_PhysRevX.14.041036} and to the experiments which exhibit two channel Kondo quantum criticality followed by a  more singular rise of the response at low temperatures.  
 

\section{Appendix}

\subsection{Magnetic impurities in a Majorana  spin liquid with a Fermi surface}

 A  Majorana metal is a  spin liquid   where elementary excitations are  Majorana neutral  fermions with a Fermi surface. This situation is realized in an exactly soluble  spin model,  which  generalizes the famous Kitaev model of Z$_2$ spin liquid \cite{kitaev} - the Yao-Lee (YL) model and has been investigated in several lattices \cite{hermanns_physics_2018,obrien_classification_2016,CPT1}.

 We start the discussion with a situation when a spin $\vec\s$ on one site of the hyperoctagon lattice is connected with an additional spin 1/2 ${\bf S}$ which couples locally to the spin degree of freedom of the Yao Lee model,  via an antiferromagnetic exchange interaction.
 
 We assume that the temperature $T << T_{c}$ so that  the visons are rare and can be neglected.  In this temperature range the YL spin liquid acts as a thermal metal: a Majorana- neutral Fermi liquid with well defined quasiparticles near the Fermi surface. Taking into account that the bulk spin operator is bilinear in Majorana fermions (\ref{spinYL}) we obtain the following single impurity Hamiltonian describing the coupling of the impurity spin to the low energy degrees of freedom of the Yao Lee model:  
\bea
H_{ex} = \frac{\ri}{2} J_K \Big({\bf S}[\vec\chi\times\vec\chi]_{x=0}\Big) + \sum_k \epsilon(k)\Big(\vec\chi(-k)\vec\chi(k)\Big), \label{MajKondo}
\eea
Since the commutation relations of the Majorana current operator coincide with the SU$_2$(2) Kac-Moody current, the dynamics of the impurity spin  is the same as in the 2-channel Kondo model (see \cite{Ioffe}). The exact solution can be generalized on the case of anisotropic exchange interaction as, for example, in \cite{AnKondo4}. Alternatively, the solution or the U(1) symmetric case $J_x= J_y = J_{\perp} \neq J_z$  can be obtained by means of bosonization as in \cite{AnKondo1}.  
 
In the standard (Dirac) formulation of the $k$-channel Kondo model, such as the one described above,  the exchange interaction includes $\sum_{m=1}^k\psi^+_{m\alpha}\s^a_{\alpha\beta}\psi_{m\beta}$ combination which is SU$_k$(2) Kac-Moody current. The anisotropy in the spin space does not introduce  currents with different symmetry and so  is irrelevant. In contrast to that  a channel anisotropy does  introduce coupling to the currents with other symmetry and this destabilizes the critical point. The contrast between the Dirac and Majorana version (\ref{MajKondo}) of the 2-channel Kondo model is that in the latter case there is no analogue of the channel anisotropy. So the QCP in the Majorana version  is robust.  If the YL model is robust, the QCP is robust too.

\bea
H_{YL}=K/2\sum_{<i,j>}\lambda^{\alpha_{ij}}_i\lambda^{\alpha_{ij}}_j({\vec\s}_i{\vec \s}_j), \label{CPTComp}
\eea
  Each site of a lattice with coordination number three contains two kinds of spins 1/2 described by $\s^a$ and $\lambda^{\alpha}$  operators respectively.  $\s^a$ operators are components of spin S=1/2 and $\lambda^{\alpha}$ are Pauli matrices describing the orbital degrees of freedom.  

 The anisotropic Ising coupling between orbitals induces Majorana fractionalization \cite{YL}
		of spins and orbitals: 
		\be
		\vec{\s}_{j}=-
		\frac{i}{2}[\vec{\chi}_j\times\vec{\chi}_j], ~~\vec{\lambda}_i=i \vec{b}_j\times
		\vec{b}_j \label{spinYL}
		\ee
        In the physical Hilbert space, where
		$\s^a_j\lambda^{\alpha}_j=2i \chi^a_j b^{\alpha}_j$, the
		fractionalized form of the Yao-Lee Hamiltonian
		$H_{YL}$ (\ref{CPTComp}) is 
		\begin{equation}
			H_{YL}=\ri K\sum_{<i,j>}\hat{u}_{ij}(\vec{\chi}_i\cdot \vec{\chi}_{j}).
		\end{equation}
		Here, $\hat{u}_{ij}=i b^{\alpha_{ij}}_ib^{\alpha_{ij}}_j$ are the  static
		$\mathbb{Z}_2$ gauge fields, (i.e. $[H_{YL},\hat{u}_{ij}]=0$).

        In three dimensions, $\mathbb{Z}_{2}$ gauge theories undergo a finite
		temperature Ising phase transition at $T_{c}$, into a deconfined phase, in
		which the visons (plaquettes with a $\pi$ flux) are
		linearly confined, and the low energy elementary excitations are a triplet of free Majorana fermions.

\begin{figure}[t]
\includegraphics[width=0.2 \textwidth]{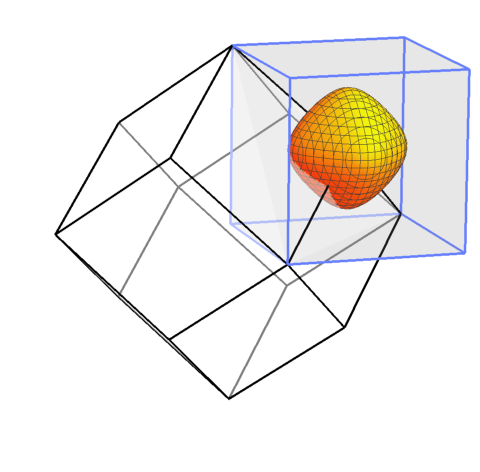}
\vspace{-0.2cm}
\caption{Majorana Fermi surface of the YL model on hyperoctagon lattice. Adopted from paper \cite{CPT1}. }
\label{Fig:YL}
\vspace{-0.5cm}
\end{figure}


\subsection{Bethe ansatz for $N =3$. T=0 results}

 The exact solution of model (12) of the main text is possible in the limit $T_{RKKY}/T_K \rightarrow 0$ when one can neglect the first term. In that limit model (12) with $ ¯\tilde I_1 =  \tilde I_2 < \tilde I_3$ is equivalent to the XXZ 6-channel Kondo model.   The Bethe ansatz equations for the XXZ 6-channel Kondo model are 
 \bea
  && [e_6(\lambda_a)]^L e_1(\lambda_a -1/g) = \prod_{b\neq a}^M e_2(\lambda_a - \lambda_b), \nonumber\\
  && E = \frac{1}{2\ri}\sum_a^M \ln[e_6(\lambda_a)] - h(6L -M), \\
  && e_n(x) = \frac{\sinh[\gamma(x - \ri n/2)]}{\sinh[\gamma(x+\ri n/2)]},\nonumber
  \eea
  where $\gamma$ is a parameter related to the exchange anisotropy. In the isotropic limit $\gamma \rightarrow 0$. These are bare equations and one has to perform some standard manipulations to derive the Thermodynamic Bethe ansatz equations describing the thermodynamic limit. 

   Our first goal is to demonstrate that the anisotropy of the exchange interaction does not affect the critical point. The easiest way to demonstrate this is to add an additional relevant term to the Hamiltonian (12) imitating the intersite tunneling of the Majorana zero modes:
   \bea
   \delta H = \ri{\cal H}\xi_1\xi_2,
   \eea
which plays the role of magnetic field and study the behavior of "magnetization" 
\bea
{\cal M} = - \frac{\p F_{imp}}{\p{\cal H}},
\eea
at $T=0$. The result will show that the power law dependence of ${\cal M}$ on ${\cal H}$ at ${\cal H} < T_{RKKY}$ is independent on the anisotropy.

We consider the simplest case $\gamma = \pi/\nu$, where $\nu > 6$ is an integer. The equations for  the ground state is
\bea
&& \rho(u) + \int_B^{\infty} K_{\gamma}(u-v)\tilde\rho(v) \rd v = \nonumber\\
&&\re^{-\pi u} \re^{-\pi B} + \frac{1}{L} f[u-1/\pi \ln(\Lambda/T_{RKKY})], \label{WH} \\
&& \epsilon^{(-)}(u) + \int_B^{\infty} K_{\gamma}(u-v)\epsilon^{(+)}(v) \rd v = - \Lambda\re^{-\pi u} \re^{-\pi B} \nonumber\\
&&+{\cal H},\label{WH2}\\
&& K_{\gamma}(\omega) = \frac{\tanh(\omega/2) \sinh (\nu\omega/2)}{2 \sinh[(\nu - 6)\omega/2]\sinh (3\omega)}, ~~ \nu = \pi/\gamma\nonumber\\
&& f(\omega) = \frac{\tanh(\omega/2)}{2\sinh(3\omega)}. 
\eea
 where $\rho, \tilde\rho$ are densities of rapidities in the $k =6$-th string and holes respectively and $\epsilon^{(-)} <0, ~~ \epsilon^{(+)} >0$ are negative and positive parts of the dispersion. 
 
 The "magnetic moment" ${\cal M} = - \p F/\p H$ is 
 \bea
 {\cal M} = \int_B^{\infty}\tilde\rho(u)\rd u.
 \eea
 The integration limit $B$ is determined by the condition that the bulk magnetization is ${\cal M}_{bulk} = {\cal H}/\Lambda$ or equivalently by the condition $\epsilon(B) =0$. Equations (\ref{WH},\ref{WH2}) are the Wiener-Hopf ones and can be solved analytically as, for example, in \cite{TsvW}. The result for the magnetic moment is 
 \begin{widetext}
\bea
  && {\cal M} = \ri \frac{G^{(+)}(0)}{2\pi} \int_{-\infty}^{\infty} \frac{\rd\omega G^{(-)}(\omega)\tanh(\omega/2)}{(\omega - \ri 0)\sinh(3\omega)}\exp\Big[\ri\frac{\omega}{\pi}\ln({\cal H}/T_{RKKY})\Big],\label{M}
\eea
\end{widetext}
  where 
 \bea
 &&K_{\gamma}(\omega) = [G^{(+)}(\omega)G^{(-)}(\omega)]^{-1},\\
 && G^{(-)}(\omega) = G^{(+)}(-\omega) = \\
 &&\frac{[12\pi(\nu -6)/\nu]^{1/2}\Gamma(1 +\ri\omega/2\pi)\Gamma(1 + \ri\nu\omega/2\pi)}{\Gamma(1/2 +\ri\omega/2\pi)\Gamma(1 + 3\ri\omega/\pi)\Gamma(1 +\ri(\nu -6)\omega/2\pi)},\nonumber
 \eea
 where $G{(\pm)}(\omega)$ are analytic in upper (lower) $\omega$-plane respectively.
 The crossover from weak coupling to strong coupling regime occurs around $H= T_{RKKY}$. When $H < T_{RKKY}$ one can bend the integration contour in (\ref{M}) to the lower half plane where $G^{(-)}(\omega)$ is analytic and the singularities  come from the poles  of the hyperbolic functions. Since the only function depending on the anisotropy is $G^{(-)}$, the anisotropy reveals itself only in the numerical coefficients in the expansion in 
  $
  ({\cal H}/T_{RKKY})^{n/3}, ~~ n=1,2, ... ; ~~({\cal H}/T_{RKKY}) ^{(1+2n)}, ~~ n=0,1, ...
   $
   In the leading order we have ${\cal M} \sim ({\cal H}/T_{RKKY})^{1/3}$ which corresponds to $h_{M} =1/4$. 
   
   
 \subsubsection{ Bethe ansatz for $N =3$. The thermodynamic equations}
 
 The TBA equations for the $k$-channel Kondo model with anisotropy and S=1/2 impurity are \cite{TsvMulti}:
 \bea
 && \phi_j = s*\ln(1+ \re^{\phi_{j-1}})(1+ \re^{\phi_{j+1}}) - \delta_{j,k} \re^{-\pi\lambda},\nonumber\\
 && j=1,...\nu -2 , ~~\nu = \pi/\gamma, \label{TBA0}\\
 &&\phi_{\nu} = \phi_{\nu -1} = s*\ln(1+\re^{\phi_{\nu -2}}), \nonumber
\eea
 and the free energy is
 \bea
 && F = F_{bulk} + \frac{1}{L}F_{imp},\\
 && F_{bulk} = -T^2\int \rd\lambda \re^{-\pi\lambda}\ln(1+ \re^{\phi_k(\lambda)}),\\
 && F_{imp} = \nonumber\\
 && -T^2\int \rd\lambda s[\lambda - \frac{1}{\pi}\ln(T/T_{RKKY})]\ln(1+\re^{\phi_1(\lambda)}).
 \eea

\subsection{Multichannel Kondo lattice}
The action of  the  multichannel Kondo lattice  is given by:

\begin{equation}
S_{imp} = S_{\text{conduction}} + S_{\text{impurity}} + S_{\text{interaction}} \label{eq:action}
\end{equation}
where  ~$\mathbf{s}_j(0) = \psi_{jm}^\dagger \frac{\vec{\sigma}}{2} \psi_{jm}, \qquad   
    \mathbf{S} = f^\dagger \frac{\vec{\sigma}}{2} f,$  and
    
\begin{widetext}
\bea
   && S_{imp} = S_{\text{conduction}} + S_{\text{impurity}} + S_{\text{interaction}},  \label{eq:action} \\
   && S_{\text{conduction}} = \sum_{m=1}^2 \sum_{\sigma=1}^2 \sum_{i=1}^N \sum_{j=1}^N \int d\tau \int d\tau' \, 
    \psi_{mj\sigma}^\dagger (\partial_\tau  +\Delta_{i,j}(\tau-\tau')) \psi_{mj\sigma}, \\
   && S_{\text{impurity}} =\sum_{\sigma=1}^2  \int d\tau \,  f_\sigma^\dagger (\partial_\tau + \lambda) f_\sigma -\lambda, \\
   && S_{\text{interaction}} = J\sum_{m=1}^2  \sum_{j=1}^N \int d\tau \, \mathbf{S} \cdot \mathbf{s}_{jm}(0)
\eea
\end{widetext}

$\Delta_{i,j}  $  is a matrix in cluster site indices i  j,  and is  diagonal in  channel indices.  It is determined by the cluster DMFT  self consistency condition.  To demonstrate   the relevance of the  new fixed point, we use   the Dynamical Cluster Approximation  (DCA) \cite{DCA_PhysRevB.58.R7475} which makes the hybridization matrix cyclical,   and its Fourier transform diagonal in the cluster momenta $K_i$. This makes  it straightforward to identify the channels, which are now labeled by the cluster momenta.  The DCA can also be understood in terms of a coarse graining of the lattice Brillouin zone into N patches of  the same size, with the cluster momenta  $K_i$ labeling each patch.

\begin{widetext}
    
\begin{equation}
\ri \omega  I - \Delta(\ri \omega,  K_i) =\Big[ \sum_{k  \in  RBZ}  \frac{1}{\ri \omega - \hat{t} _{k+K_i} - \Sigma_{imp} (\ri \omega, \ K_i)} \Big]^{-1}+ \Sigma_{imp} (\ri \omega, \ K_i)
\label{scc}
\end{equation}
\end{widetext}

Under the  assumption   that the impurity self energy $\Sigma_{imp}$  of the problem under consideration is a  not divergent at low frequency,  analytic continuation  of Eqn. \ref{scc} gives a finite  value 
$-Im \Delta(\omega + i \delta ) $  at zero frequency,   coming  from the k integration in the patches which cross the fermi level. This statement is valid   as  long as the conduction  bands are sufficiently broad so that the real part of the self energy $\Sigma_{imp}$  does not shift   zeros of the denominator in the sum over k in   Eqn. \ref{scc} beyond the edge of the band of the conduction electrons.  Having  shown that  that $-Im \Delta(\omega + i \delta ) \neq 0 $ we  now have the spin sector impurity model considered in the text,  together with a decoupled charge and channel sector, and the spin spin correlation functions of the cluster DMFT solution at low energies are given by the results presented in the  main text.  The evaluation of correlation functions involving the charge sector are left for future studies.

\textit{Acknowledgments:}
			We would like to thank Piers Coleman and Andriy Nevidomskyy for fruitful discussions. This work was supported by Office of Basic Energy Sciences, Material
		Sciences and Engineering Division, U.S. Department of Energy (DOE)
		under Contracts No. DE-SC0012704 and by the National Science Foundation (GK)
under DMR-1733071.


\bibliography{sample}

\begin{thebibliography}{45}%
\makeatletter
\providecommand \@ifxundefined [1]{%
 \@ifx{#1\undefined}
}%
\providecommand \@ifnum [1]{%
 \ifnum #1\expandafter \@firstoftwo
 \else \expandafter \@secondoftwo
 \fi
}%
\providecommand \@ifx [1]{%
 \ifx #1\expandafter \@firstoftwo
 \else \expandafter \@secondoftwo
 \fi
}%
\providecommand \natexlab [1]{#1}%
\providecommand \enquote  [1]{``#1''}%
\providecommand \bibnamefont  [1]{#1}%
\providecommand \bibfnamefont [1]{#1}%
\providecommand \citenamefont [1]{#1}%
\providecommand \href@noop [0]{\@secondoftwo}%
\providecommand \href [0]{\begingroup \@sanitize@url \@href}%
\providecommand \@href[1]{\@@startlink{#1}\@@href}%
\providecommand \@@href[1]{\endgroup#1\@@endlink}%
\providecommand \@sanitize@url [0]{\catcode `\\12\catcode `\$12\catcode `\&12\catcode `\#12\catcode `\^12\catcode `\_12\catcode `\%12\relax}%
\providecommand \@@startlink[1]{}%
\providecommand \@@endlink[0]{}%
\providecommand \url  [0]{\begingroup\@sanitize@url \@url }%
\providecommand \@url [1]{\endgroup\@href {#1}{\urlprefix }}%
\providecommand \urlprefix  [0]{URL }%
\providecommand \Eprint [0]{\href }%
\providecommand \doibase [0]{https://doi.org/}%
\providecommand \selectlanguage [0]{\@gobble}%
\providecommand \bibinfo  [0]{\@secondoftwo}%
\providecommand \bibfield  [0]{\@secondoftwo}%
\providecommand \translation [1]{[#1]}%
\providecommand \BibitemOpen [0]{}%
\providecommand \bibitemStop [0]{}%
\providecommand \bibitemNoStop [0]{.\EOS\space}%
\providecommand \EOS [0]{\spacefactor3000\relax}%
\providecommand \BibitemShut  [1]{\csname bibitem#1\endcsname}%
\let\auto@bib@innerbib\@empty
\bibitem [{\citenamefont {Hewson}(1993)}]{Hewson1993}%
  \BibitemOpen
  \bibfield  {author} {\bibinfo {author} {\bibfnamefont {A.~C.}\ \bibnamefont {Hewson}},\ }\href {https://doi.org/10.1017/CBO9780511470752} {\emph {\bibinfo {title} {The Kondo Problem to Heavy Fermions}}},\ \bibinfo {series} {Cambridge Studies in Magnetism}, Vol.~\bibinfo {volume} {2}\ (\bibinfo  {publisher} {Cambridge University Press},\ \bibinfo {address} {Cambridge, UK},\ \bibinfo {year} {1993})\BibitemShut {NoStop}%
\bibitem [{\citenamefont {Varma}\ \emph {et~al.}(1989)\citenamefont {Varma}, \citenamefont {Littlewood}, \citenamefont {Schmitt-Rink}, \citenamefont {Abrahams},\ and\ \citenamefont {Ruckenstein}}]{PhysRevLett.63.1996}%
  \BibitemOpen
  \bibfield  {author} {\bibinfo {author} {\bibfnamefont {C.~M.}\ \bibnamefont {Varma}}, \bibinfo {author} {\bibfnamefont {P.~B.}\ \bibnamefont {Littlewood}}, \bibinfo {author} {\bibfnamefont {S.}~\bibnamefont {Schmitt-Rink}}, \bibinfo {author} {\bibfnamefont {E.}~\bibnamefont {Abrahams}},\ and\ \bibinfo {author} {\bibfnamefont {A.~E.}\ \bibnamefont {Ruckenstein}},\ }\bibfield  {title} {\bibinfo {title} {Phenomenology of the normal state of cu-o high-temperature superconductors},\ }\href {https://doi.org/10.1103/PhysRevLett.63.1996} {\bibfield  {journal} {\bibinfo  {journal} {Phys. Rev. Lett.}\ }\textbf {\bibinfo {volume} {63}},\ \bibinfo {pages} {1996} (\bibinfo {year} {1989})}\BibitemShut {NoStop}%
\bibitem [{\citenamefont {Si}\ \emph {et~al.}(2001)\citenamefont {Si}, \citenamefont {Rabello}, \citenamefont {Ingersent},\ and\ \citenamefont {Smith}}]{Si2001}%
  \BibitemOpen
  \bibfield  {author} {\bibinfo {author} {\bibfnamefont {Q.}~\bibnamefont {Si}}, \bibinfo {author} {\bibfnamefont {S.}~\bibnamefont {Rabello}}, \bibinfo {author} {\bibfnamefont {K.}~\bibnamefont {Ingersent}},\ and\ \bibinfo {author} {\bibfnamefont {J.~L.}\ \bibnamefont {Smith}},\ }\bibfield  {title} {\bibinfo {title} {Locally critical quantum phase transitions in strongly correlated metals},\ }\href {https://doi.org/10.1038/35101507} {\bibfield  {journal} {\bibinfo  {journal} {Nature}\ }\textbf {\bibinfo {volume} {413}},\ \bibinfo {pages} {804} (\bibinfo {year} {2001})}\BibitemShut {NoStop}%
\bibitem [{\citenamefont {Oeschler}\ \emph {et~al.}(2008)\citenamefont {Oeschler}, \citenamefont {Hartmann}, \citenamefont {Pikul}, \citenamefont {Krellner}, \citenamefont {Geibel},\ and\ \citenamefont {Steglich}}]{OESCHLER20081254}%
  \BibitemOpen
  \bibfield  {author} {\bibinfo {author} {\bibfnamefont {N.}~\bibnamefont {Oeschler}}, \bibinfo {author} {\bibfnamefont {S.}~\bibnamefont {Hartmann}}, \bibinfo {author} {\bibfnamefont {A.}~\bibnamefont {Pikul}}, \bibinfo {author} {\bibfnamefont {C.}~\bibnamefont {Krellner}}, \bibinfo {author} {\bibfnamefont {C.}~\bibnamefont {Geibel}},\ and\ \bibinfo {author} {\bibfnamefont {F.}~\bibnamefont {Steglich}},\ }\bibfield  {title} {\bibinfo {title} {Low-temperature specific heat of ybrh2si2},\ }\href {https://doi.org/https://doi.org/10.1016/j.physb.2007.10.119} {\bibfield  {journal} {\bibinfo  {journal} {Physica B: Condensed Matter}\ }\textbf {\bibinfo {volume} {403}},\ \bibinfo {pages} {1254} (\bibinfo {year} {2008})}\BibitemShut {NoStop}%
\bibitem [{\citenamefont {Guo}\ \emph {et~al.}(2024)\citenamefont {Guo}, \citenamefont {Chen}, \citenamefont {Hoveyda-Marashi}, \citenamefont {Bettler}, \citenamefont {Chaudhuri}, \citenamefont {Kengle}, \citenamefont {Schneeloch}, \citenamefont {Zhang}, \citenamefont {Gu}, \citenamefont {Chiang}, \citenamefont {Tsvelik}, \citenamefont {Faulkner}, \citenamefont {Phillips},\ and\ \citenamefont {Abbamonte}}]{CFTguo2024conformallyinvariantchargefluctuations}%
  \BibitemOpen
  \bibfield  {author} {\bibinfo {author} {\bibfnamefont {X.}~\bibnamefont {Guo}}, \bibinfo {author} {\bibfnamefont {J.}~\bibnamefont {Chen}}, \bibinfo {author} {\bibfnamefont {F.}~\bibnamefont {Hoveyda-Marashi}}, \bibinfo {author} {\bibfnamefont {S.~L.}\ \bibnamefont {Bettler}}, \bibinfo {author} {\bibfnamefont {D.}~\bibnamefont {Chaudhuri}}, \bibinfo {author} {\bibfnamefont {C.~S.}\ \bibnamefont {Kengle}}, \bibinfo {author} {\bibfnamefont {J.~A.}\ \bibnamefont {Schneeloch}}, \bibinfo {author} {\bibfnamefont {R.}~\bibnamefont {Zhang}}, \bibinfo {author} {\bibfnamefont {G.}~\bibnamefont {Gu}}, \bibinfo {author} {\bibfnamefont {T.-C.}\ \bibnamefont {Chiang}}, \bibinfo {author} {\bibfnamefont {A.~M.}\ \bibnamefont {Tsvelik}}, \bibinfo {author} {\bibfnamefont {T.}~\bibnamefont {Faulkner}}, \bibinfo {author} {\bibfnamefont {P.~W.}\ \bibnamefont {Phillips}},\ and\ \bibinfo {author} {\bibfnamefont {P.}~\bibnamefont {Abbamonte}},\ }\bibfield  {title} {\bibinfo {title} {Conformally invariant charge fluctuations in a
  strange metal},\ }\href {https://arxiv.org/abs/2411.11164} {\bibfield  {journal} {\bibinfo  {journal} {arXiv preprint}\ } (\bibinfo {year} {2024})},\ \Eprint {https://arxiv.org/abs/2411.11164} {arXiv:2411.11164 [cond-mat.str-el]} \BibitemShut {NoStop}%
\bibitem [{\citenamefont {Craig}\ \emph {et~al.}(2011)\citenamefont {Craig}, \citenamefont {Taylor}, \citenamefont {Lester}, \citenamefont {Marcus}, \citenamefont {Hanson},\ and\ \citenamefont {Gossard}}]{Experiment1}%
  \BibitemOpen
  \bibfield  {author} {\bibinfo {author} {\bibfnamefont {N.~J.}\ \bibnamefont {Craig}}, \bibinfo {author} {\bibfnamefont {J.~M.}\ \bibnamefont {Taylor}}, \bibinfo {author} {\bibfnamefont {E.}~\bibnamefont {Lester}}, \bibinfo {author} {\bibfnamefont {C.~M.}\ \bibnamefont {Marcus}}, \bibinfo {author} {\bibfnamefont {M.}~\bibnamefont {Hanson}},\ and\ \bibinfo {author} {\bibfnamefont {A.~C.}\ \bibnamefont {Gossard}},\ }\bibfield  {title} {\bibinfo {title} {Tunable nonlocal spin control in a coupled-quantum dot system},\ }\href {https://doi.org/10.1126/science.1095452} {\bibfield  {journal} {\bibinfo  {journal} {Science}\ }\textbf {\bibinfo {volume} {7}},\ \bibinfo {pages} {901} (\bibinfo {year} {2011})}\BibitemShut {NoStop}%
\bibitem [{\citenamefont {Bork}\ \emph {et~al.}(2011)\citenamefont {Bork}, \citenamefont {Zhang}, \citenamefont {Diekh\"oner}, \citenamefont {Borda}, \citenamefont {Simon}, \citenamefont {Kroha}, \citenamefont {Wahl},\ and\ \citenamefont {Kern}}]{Experiment2}%
  \BibitemOpen
  \bibfield  {author} {\bibinfo {author} {\bibfnamefont {J.}~\bibnamefont {Bork}}, \bibinfo {author} {\bibfnamefont {Y.-h.}\ \bibnamefont {Zhang}}, \bibinfo {author} {\bibfnamefont {L.}~\bibnamefont {Diekh\"oner}}, \bibinfo {author} {\bibfnamefont {L.}~\bibnamefont {Borda}}, \bibinfo {author} {\bibfnamefont {P.}~\bibnamefont {Simon}}, \bibinfo {author} {\bibfnamefont {J.}~\bibnamefont {Kroha}}, \bibinfo {author} {\bibfnamefont {P.}~\bibnamefont {Wahl}},\ and\ \bibinfo {author} {\bibfnamefont {K.}~\bibnamefont {Kern}},\ }\bibfield  {title} {\bibinfo {title} {A tunable two-impurity kondo system in an atomic point contact},\ }\href {https://doi.org/10.1038/nphys2076} {\bibfield  {journal} {\bibinfo  {journal} {Nat. Phys.}\ }\textbf {\bibinfo {volume} {7}},\ \bibinfo {pages} {901} (\bibinfo {year} {2011})}\BibitemShut {NoStop}%
\bibitem [{\citenamefont {Spinelli}\ \emph {et~al.}(2015)\citenamefont {Spinelli}, \citenamefont {Gerrits}, \citenamefont {Toscovic}, \citenamefont {Bryant}, \citenamefont {Ternes},\ and\ \citenamefont {Otte}}]{Spinelli}%
  \BibitemOpen
  \bibfield  {author} {\bibinfo {author} {\bibfnamefont {A.}~\bibnamefont {Spinelli}}, \bibinfo {author} {\bibfnamefont {M.}~\bibnamefont {Gerrits}}, \bibinfo {author} {\bibfnamefont {R.}~\bibnamefont {Toscovic}}, \bibinfo {author} {\bibfnamefont {B.}~\bibnamefont {Bryant}}, \bibinfo {author} {\bibfnamefont {M.}~\bibnamefont {Ternes}},\ and\ \bibinfo {author} {\bibfnamefont {A.~F.}\ \bibnamefont {Otte}},\ }\bibfield  {title} {\bibinfo {title} {Exploring the phase diagram of the two-impurity kondo problem},\ }\href {https://doi.org/10.1038/ncomm10046} {\bibfield  {journal} {\bibinfo  {journal} {Nat. Comm.}\ }\textbf {\bibinfo {volume} {6}},\ \bibinfo {pages} {10046} (\bibinfo {year} {2015})}\BibitemShut {NoStop}%
\bibitem [{\citenamefont {Keller}\ \emph {et~al.}(2015)\citenamefont {Keller}, \citenamefont {Peeters}, \citenamefont {Moca}, \citenamefont {Weymann}, \citenamefont {Mahalu}, \citenamefont {Umansky},\ and\ \citenamefont {Goldhaber-Gordon}}]{Goldhaber}%
  \BibitemOpen
  \bibfield  {author} {\bibinfo {author} {\bibfnamefont {A.}~\bibnamefont {Keller}}, \bibinfo {author} {\bibfnamefont {L.}~\bibnamefont {Peeters}}, \bibinfo {author} {\bibfnamefont {C.~P.}\ \bibnamefont {Moca}}, \bibinfo {author} {\bibfnamefont {I.}~\bibnamefont {Weymann}}, \bibinfo {author} {\bibfnamefont {D.}~\bibnamefont {Mahalu}}, \bibinfo {author} {\bibfnamefont {V.~G.}\ \bibnamefont {Umansky}, \bibfnamefont {Zar\'and}},\ and\ \bibinfo {author} {\bibfnamefont {D.}~\bibnamefont {Goldhaber-Gordon}},\ }\bibfield  {title} {\bibinfo {title} {Universal fermi liquid crossover and quantum criticality in a mesoscopic system},\ }\href {https://doi.org/10.1038/nature15261} {\bibfield  {journal} {\bibinfo  {journal} {Nature}\ }\textbf {\bibinfo {volume} {526}},\ \bibinfo {pages} {237} (\bibinfo {year} {2015})}\BibitemShut {NoStop}%
\bibitem [{\citenamefont {Eickhoff}\ and\ \citenamefont {Anders}(2021)}]{Eickhoff2021}%
  \BibitemOpen
  \bibfield  {author} {\bibinfo {author} {\bibfnamefont {F.}~\bibnamefont {Eickhoff}}\ and\ \bibinfo {author} {\bibfnamefont {F.~B.}\ \bibnamefont {Anders}},\ }\bibfield  {title} {\bibinfo {title} {Spectral properties of strongly correlated multi-impurity models in the kondo insulator regime: Emergent coherence, metallic surface states, and quantum phase transitions},\ }\href {https://doi.org/10.1103/PhysRevB.104.165105} {\bibfield  {journal} {\bibinfo  {journal} {Phys. Rev. B}\ }\textbf {\bibinfo {volume} {104}},\ \bibinfo {pages} {165105} (\bibinfo {year} {2021})}\BibitemShut {NoStop}%
\bibitem [{\citenamefont {Jones}\ and\ \citenamefont {Varma}(1989)}]{Varma}%
  \BibitemOpen
  \bibfield  {author} {\bibinfo {author} {\bibfnamefont {B.}~\bibnamefont {Jones}}\ and\ \bibinfo {author} {\bibfnamefont {C.}~\bibnamefont {Varma}},\ }\bibfield  {title} {\bibinfo {title} {Critical point in the solution of the 2 magnetic impurity problem},\ }\href {https://doi.org/10.1103/PhysRevB.40.324} {\bibfield  {journal} {\bibinfo  {journal} {Phys. Rev. B}\ }\textbf {\bibinfo {volume} {40}},\ \bibinfo {pages} {324} (\bibinfo {year} {1989})}\BibitemShut {NoStop}%
\bibitem [{\citenamefont {Georges}\ and\ \citenamefont {Sengupta}(1995)}]{georges}%
  \BibitemOpen
  \bibfield  {author} {\bibinfo {author} {\bibfnamefont {A.}~\bibnamefont {Georges}}\ and\ \bibinfo {author} {\bibfnamefont {A.~M.}\ \bibnamefont {Sengupta}},\ }\bibfield  {title} {\bibinfo {title} {Solution of the two-impuirity, two-channel kondo model},\ }\href {https://doi.org/10.1103/PhysRevLett.74.2808} {\bibfield  {journal} {\bibinfo  {journal} {Phys. Rev. Lett.}\ }\textbf {\bibinfo {volume} {74}},\ \bibinfo {pages} {2808} (\bibinfo {year} {1995})}\BibitemShut {NoStop}%
\bibitem [{\citenamefont {Georges}\ and\ \citenamefont {Sengupta}(1994)}]{AnKondo1}%
  \BibitemOpen
  \bibfield  {author} {\bibinfo {author} {\bibfnamefont {A.}~\bibnamefont {Georges}}\ and\ \bibinfo {author} {\bibfnamefont {A.~M.}\ \bibnamefont {Sengupta}},\ }\bibfield  {title} {\bibinfo {title} {Solution of the two-impuirity, two-channel kondo model},\ }\href {https://doi.org/10.1103/PhysRevLett.49.10020} {\bibfield  {journal} {\bibinfo  {journal} {Phys. Rev. Lett.}\ }\textbf {\bibinfo {volume} {49}},\ \bibinfo {pages} {10020} (\bibinfo {year} {1994})}\BibitemShut {NoStop}%
\bibitem [{\citenamefont {Ingersent}\ \emph {et~al.}(2005)\citenamefont {Ingersent}, \citenamefont {Ludwig},\ and\ \citenamefont {Affleck}}]{PhysRevLett.95.257204}%
  \BibitemOpen
  \bibfield  {author} {\bibinfo {author} {\bibfnamefont {K.}~\bibnamefont {Ingersent}}, \bibinfo {author} {\bibfnamefont {A.~W.~W.}\ \bibnamefont {Ludwig}},\ and\ \bibinfo {author} {\bibfnamefont {I.}~\bibnamefont {Affleck}},\ }\bibfield  {title} {\bibinfo {title} {Kondo screening in a magnetically frustrated nanostructure: Exact results on a stable non-fermi-liquid phase},\ }\href {https://doi.org/10.1103/PhysRevLett.95.257204} {\bibfield  {journal} {\bibinfo  {journal} {Phys. Rev. Lett.}\ }\textbf {\bibinfo {volume} {95}},\ \bibinfo {pages} {257204} (\bibinfo {year} {2005})}\BibitemShut {NoStop}%
\bibitem [{\citenamefont {Beri}\ and\ \citenamefont {Cooper}(2012)}]{BeriCooper}%
  \BibitemOpen
  \bibfield  {author} {\bibinfo {author} {\bibfnamefont {B.}~\bibnamefont {Beri}}\ and\ \bibinfo {author} {\bibfnamefont {N.}~\bibnamefont {Cooper}},\ }\bibfield  {title} {\bibinfo {title} {Topological kondo effect with majorana fermions},\ }\href {https://doi.org/10.1103/PhysRevLett.109.156803} {\bibfield  {journal} {\bibinfo  {journal} {Phys. Rev. Lett.}\ }\textbf {\bibinfo {volume} {109}},\ \bibinfo {pages} {156803} (\bibinfo {year} {2012})}\BibitemShut {NoStop}%
\bibitem [{\citenamefont {Altland}\ and\ \citenamefont {Egger}(2013)}]{AlEg}%
  \BibitemOpen
  \bibfield  {author} {\bibinfo {author} {\bibfnamefont {A.}~\bibnamefont {Altland}}\ and\ \bibinfo {author} {\bibfnamefont {R.}~\bibnamefont {Egger}},\ }\bibfield  {title} {\bibinfo {title} {Multiterminal coulomb-majorana junction},\ }\href {https://doi.org/10.1103/PhysRevLett.110.196401} {\bibfield  {journal} {\bibinfo  {journal} {Phys. Rev. Lett.}\ }\textbf {\bibinfo {volume} {110}},\ \bibinfo {pages} {196401} (\bibinfo {year} {2013})}\BibitemShut {NoStop}%
\bibitem [{\citenamefont {Altland}\ \emph {et~al.}(2014{\natexlab{a}})\citenamefont {Altland}, \citenamefont {Beri}, \citenamefont {Egger},\ and\ \citenamefont {Tsvelik}}]{AlTsv}%
  \BibitemOpen
  \bibfield  {author} {\bibinfo {author} {\bibfnamefont {A.}~\bibnamefont {Altland}}, \bibinfo {author} {\bibfnamefont {B.}~\bibnamefont {Beri}}, \bibinfo {author} {\bibfnamefont {R.}~\bibnamefont {Egger}},\ and\ \bibinfo {author} {\bibfnamefont {A.~M.}\ \bibnamefont {Tsvelik}},\ }\bibfield  {title} {\bibinfo {title} {Multichannel kondo impurity dynamics in a majorana device},\ }\href {https://doi.org/10.1103/PhysRevLett.113.076401} {\bibfield  {journal} {\bibinfo  {journal} {Phys. Rev. Lett.}\ }\textbf {\bibinfo {volume} {113}},\ \bibinfo {pages} {076401} (\bibinfo {year} {2014}{\natexlab{a}})}\BibitemShut {NoStop}%
\bibitem [{\citenamefont {Altland}\ \emph {et~al.}(2014{\natexlab{b}})\citenamefont {Altland}, \citenamefont {Beri}, \citenamefont {Egger},\ and\ \citenamefont {Tsvelik}}]{AlTsv2}%
  \BibitemOpen
  \bibfield  {author} {\bibinfo {author} {\bibfnamefont {A.}~\bibnamefont {Altland}}, \bibinfo {author} {\bibfnamefont {B.}~\bibnamefont {Beri}}, \bibinfo {author} {\bibfnamefont {R.}~\bibnamefont {Egger}},\ and\ \bibinfo {author} {\bibfnamefont {A.~M.}\ \bibnamefont {Tsvelik}},\ }\bibfield  {title} {\bibinfo {title} {Bethe ansatz solution of the topological kondo model},\ }\href {https://doi.org/10.1088/1751-8113/47/26/265001} {\bibfield  {journal} {\bibinfo  {journal} {J. Phys. A: Math. Theor.}\ }\textbf {\bibinfo {volume} {47}},\ \bibinfo {pages} {265001} (\bibinfo {year} {2014}{\natexlab{b}})}\BibitemShut {NoStop}%
\bibitem [{\citenamefont {Li}\ \emph {et~al.}(2023)\citenamefont {Li}, \citenamefont {Oreg},\ and\ \citenamefont {V\'ayrynen}}]{Jukka}%
  \BibitemOpen
  \bibfield  {author} {\bibinfo {author} {\bibfnamefont {G.}~\bibnamefont {Li}}, \bibinfo {author} {\bibfnamefont {Y.}~\bibnamefont {Oreg}},\ and\ \bibinfo {author} {\bibfnamefont {J.~I.}\ \bibnamefont {V\'ayrynen}},\ }\bibfield  {title} {\bibinfo {title} {Multichannel topological kondo effect},\ }\href {https://doi.org/10.1103/PhysRevLett.130.066302} {\bibfield  {journal} {\bibinfo  {journal} {Phys. Rev. Lett.}\ }\textbf {\bibinfo {volume} {130}},\ \bibinfo {pages} {066302} (\bibinfo {year} {2023})}\BibitemShut {NoStop}%
\bibitem [{\citenamefont {Cox}(1994)}]{cox1994resistivitytwochannelkondomodel}%
  \BibitemOpen
  \bibfield  {author} {\bibinfo {author} {\bibfnamefont {D.~L.}\ \bibnamefont {Cox}},\ }\href {https://arxiv.org/abs/cond-mat/9405049} {\bibinfo {title} {Resistivity of the two-channel kondo model in infinite dimensions}} (\bibinfo {year} {1994}),\ \Eprint {https://arxiv.org/abs/cond-mat/9405049} {arXiv:cond-mat/9405049 [cond-mat]} \BibitemShut {NoStop}%
\bibitem [{\citenamefont {Lenk}\ \emph {et~al.}(2024)\citenamefont {Lenk}, \citenamefont {Gao}, \citenamefont {Kroha},\ and\ \citenamefont {Nevidomskyy}}]{PhysRevRes6}%
  \BibitemOpen
  \bibfield  {author} {\bibinfo {author} {\bibfnamefont {M.}~\bibnamefont {Lenk}}, \bibinfo {author} {\bibfnamefont {F.}~\bibnamefont {Gao}}, \bibinfo {author} {\bibfnamefont {J.}~\bibnamefont {Kroha}},\ and\ \bibinfo {author} {\bibfnamefont {H.~A.}\ \bibnamefont {Nevidomskyy}},\ }\bibfield  {title} {\bibinfo {title} {Strange-metal behavior without fine-tuning in prv$_2$al$_{20}$},\ }\href {https://doi.org/10.1103/PhysRevRes.6.L042008} {\bibfield  {journal} {\bibinfo  {journal} {Phys. Rev. Res}\ }\textbf {\bibinfo {volume} {6}},\ \bibinfo {pages} {L042008} (\bibinfo {year} {2024})}\BibitemShut {NoStop}%
\bibitem [{\citenamefont {Sakai}\ and\ \citenamefont {Nakatsuji}(2011)}]{Nakatsuji}%
  \BibitemOpen
  \bibfield  {author} {\bibinfo {author} {\bibfnamefont {A.}~\bibnamefont {Sakai}}\ and\ \bibinfo {author} {\bibfnamefont {S.}~\bibnamefont {Nakatsuji}},\ }\bibfield  {title} {\bibinfo {title} {Kondo effects and multipolar order in the cubic prtr$_2$al$_{20}$ (tr=ti, v)},\ }\href {https://doi.org/10.1143/JPSJ.80.063701} {\bibfield  {journal} {\bibinfo  {journal} {Journal of the Physical Society of Japan}\ }\textbf {\bibinfo {volume} {80}},\ \bibinfo {pages} {063701} (\bibinfo {year} {2011})},\ \Eprint {https://arxiv.org/abs/https://doi.org/10.1143/JPSJ.80.063701} {https://doi.org/10.1143/JPSJ.80.063701} \BibitemShut {NoStop}%
\bibitem [{\citenamefont {Coleman}\ \emph {et~al.}(1995)\citenamefont {Coleman}, \citenamefont {Ioffe},\ and\ \citenamefont {Tsvelik}}]{Ioffe}%
  \BibitemOpen
  \bibfield  {author} {\bibinfo {author} {\bibfnamefont {P.}~\bibnamefont {Coleman}}, \bibinfo {author} {\bibfnamefont {L.~B.}\ \bibnamefont {Ioffe}},\ and\ \bibinfo {author} {\bibfnamefont {A.~M.}\ \bibnamefont {Tsvelik}},\ }\bibfield  {title} {\bibinfo {title} {Simple formulation of the two-channel kondo model},\ }\href {https://doi.org/10.1103/PhysRevB.52.6611} {\bibfield  {journal} {\bibinfo  {journal} {Phys. Rev. B}\ }\textbf {\bibinfo {volume} {52}},\ \bibinfo {pages} {6611} (\bibinfo {year} {1995})}\BibitemShut {NoStop}%
\bibitem [{\citenamefont {Ingersent}\ \emph {et~al.}(1992)\citenamefont {Ingersent}, \citenamefont {Jones},\ and\ \citenamefont {Wilkins}}]{ingersent}%
  \BibitemOpen
  \bibfield  {author} {\bibinfo {author} {\bibfnamefont {K.}~\bibnamefont {Ingersent}}, \bibinfo {author} {\bibfnamefont {B.~A.}\ \bibnamefont {Jones}},\ and\ \bibinfo {author} {\bibfnamefont {J.~W.}\ \bibnamefont {Wilkins}},\ }\bibfield  {title} {\bibinfo {title} {Study of the two-impurity, two-channel kondo hamiltonian},\ }\href {https://doi.org/10.1103/PhysRevLett.69.2594} {\bibfield  {journal} {\bibinfo  {journal} {Phys. Rev. Lett.}\ }\textbf {\bibinfo {volume} {69}},\ \bibinfo {pages} {2594} (\bibinfo {year} {1992})}\BibitemShut {NoStop}%
\bibitem [{\citenamefont {Kirillov}\ and\ \citenamefont {Reshetikhin}(1987)}]{AnKondo2}%
  \BibitemOpen
  \bibfield  {author} {\bibinfo {author} {\bibfnamefont {A.~N.}\ \bibnamefont {Kirillov}}\ and\ \bibinfo {author} {\bibfnamefont {N.~Y.}\ \bibnamefont {Reshetikhin}},\ }\bibfield  {title} {\bibinfo {title} {Exact solution of the integrable xxz heisenberg model with arbitrary spin: I. the ground state and the excitation spectrum},\ }\href {https://doi.org/10.1088/0305-4470/20/6/038} {\bibfield  {journal} {\bibinfo  {journal} {J. Phys. A: Math. Gen.}\ }\textbf {\bibinfo {volume} {20}},\ \bibinfo {pages} {1565} (\bibinfo {year} {1987})}\BibitemShut {NoStop}%
\bibitem [{\citenamefont {Kirillov}\ and\ \citenamefont {Reshetikhin}(1985)}]{AnKondo3}%
  \BibitemOpen
  \bibfield  {author} {\bibinfo {author} {\bibfnamefont {A.~N.}\ \bibnamefont {Kirillov}}\ and\ \bibinfo {author} {\bibfnamefont {N.~Y.}\ \bibnamefont {Reshetikhin}},\ }\bibfield  {title} {\bibinfo {title} {Exact solution of the integrable xxz heisenberg model with arbitrary spin: I. the ground state and the excitation spectrum},\ }\href@noop {} {\bibfield  {journal} {\bibinfo  {journal} {Zap. Nauch. Seminar. LOMI}\ }\textbf {\bibinfo {volume} {145}},\ \bibinfo {pages} {109} (\bibinfo {year} {1985})}\BibitemShut {NoStop}%
\bibitem [{\citenamefont {Tsvelik}(1995)}]{AnKondo4}%
  \BibitemOpen
  \bibfield  {author} {\bibinfo {author} {\bibfnamefont {A.~M.}\ \bibnamefont {Tsvelik}},\ }\bibfield  {title} {\bibinfo {title} {Toulouse limit of the multichannel kondo model},\ }\href {https://doi.org/10.1103/PhysRevB.52.4366} {\bibfield  {journal} {\bibinfo  {journal} {Phys. Rev. B}\ }\textbf {\bibinfo {volume} {52}},\ \bibinfo {pages} {4366} (\bibinfo {year} {1995})}\BibitemShut {NoStop}%
\bibitem [{\citenamefont {Tsvelik}(1985)}]{TsvMulti}%
  \BibitemOpen
  \bibfield  {author} {\bibinfo {author} {\bibfnamefont {A.~M.}\ \bibnamefont {Tsvelik}},\ }\bibfield  {title} {\bibinfo {title} {Thermodynamics of the multi-channel kondo problem},\ }\href {https://doi.org/0.1088/0022-3719/18/1/020} {\bibfield  {journal} {\bibinfo  {journal} {J. Phys. C: Cond.Mat.}\ }\textbf {\bibinfo {volume} {18}},\ \bibinfo {pages} {159} (\bibinfo {year} {1985})}\BibitemShut {NoStop}%
\bibitem [{\citenamefont {Kimura}(2021)}]{kimura}%
  \BibitemOpen
  \bibfield  {author} {\bibinfo {author} {\bibfnamefont {T.}~\bibnamefont {Kimura}},\ }\bibfield  {title} {\bibinfo {title} {Abcd of the kondo effect},\ }\href {https://doi.org/10.7566/JPSJ.90.024708} {\bibfield  {journal} {\bibinfo  {journal} {J. Phys. Soc. Jap.}\ }\textbf {\bibinfo {volume} {90}},\ \bibinfo {pages} {024708} (\bibinfo {year} {2021})}\BibitemShut {NoStop}%
\bibitem [{\citenamefont {Kitaev}(2006)}]{kitaev}%
  \BibitemOpen
  \bibfield  {author} {\bibinfo {author} {\bibfnamefont {A.}~\bibnamefont {Kitaev}},\ }\bibfield  {title} {\bibinfo {title} {Anyons in an exactly solvable model and beyond},\ }\href {https://doi.org/10.1016/j.aop.2005.10.005} {\bibfield  {journal} {\bibinfo  {journal} {Ann. Phys.}\ }\textbf {\bibinfo {volume} {321}},\ \bibinfo {pages} {2} (\bibinfo {year} {2006})}\BibitemShut {NoStop}%
\bibitem [{\citenamefont {Hermanns}\ and\ \citenamefont {Trebst}(2014)}]{hermanns_physics_2018}%
  \BibitemOpen
  \bibfield  {author} {\bibinfo {author} {\bibfnamefont {M.}~\bibnamefont {Hermanns}}\ and\ \bibinfo {author} {\bibfnamefont {S.}~\bibnamefont {Trebst}},\ }\bibfield  {title} {\bibinfo {title} {Quantum spin liquid with a majorana fermi surface on the three-dimensional hyperoctogon lattice},\ }\href {https://doi.org/10.1103/PhysRevB.89.235102} {\bibfield  {journal} {\bibinfo  {journal} {Phys. Rev. B}\ }\textbf {\bibinfo {volume} {89}},\ \bibinfo {pages} {235102} (\bibinfo {year} {2014})}\BibitemShut {NoStop}%
\bibitem [{\citenamefont {O'Brien}\ \emph {et~al.}(2016)\citenamefont {O'Brien}, \citenamefont {Hermanns},\ and\ \citenamefont {Trebst}}]{obrien_classification_2016}%
  \BibitemOpen
  \bibfield  {author} {\bibinfo {author} {\bibfnamefont {K.}~\bibnamefont {O'Brien}}, \bibinfo {author} {\bibfnamefont {M.}~\bibnamefont {Hermanns}},\ and\ \bibinfo {author} {\bibfnamefont {S.}~\bibnamefont {Trebst}},\ }\bibfield  {title} {\bibinfo {title} {Classification of gapless z$_2$ spin liquids in three-dimensional kitaev models},\ }\href {https://doi.org/10.1103/PhysRevB.93.085101} {\bibfield  {journal} {\bibinfo  {journal} {Phys. Rev. B}\ }\textbf {\bibinfo {volume} {93}},\ \bibinfo {pages} {085101} (\bibinfo {year} {2016})}\BibitemShut {NoStop}%
\bibitem [{\citenamefont {Coleman}\ \emph {et~al.}(2022)\citenamefont {Coleman}, \citenamefont {Panigrahy},\ and\ \citenamefont {Tsvelik}}]{CPT1}%
  \BibitemOpen
  \bibfield  {author} {\bibinfo {author} {\bibfnamefont {P.}~\bibnamefont {Coleman}}, \bibinfo {author} {\bibfnamefont {A.}~\bibnamefont {Panigrahy}},\ and\ \bibinfo {author} {\bibfnamefont {A.~M.}\ \bibnamefont {Tsvelik}},\ }\bibfield  {title} {\bibinfo {title} {A solvable 3d kondo lattice exhibiting odd-frequency pairing and order fractionalization},\ }\href {https://doi.org/10.1103/PhysRevLett.129.177601} {\bibfield  {journal} {\bibinfo  {journal} {Phys. Rev. Lett.}\ }\textbf {\bibinfo {volume} {129}},\ \bibinfo {pages} {177601} (\bibinfo {year} {2022})}\BibitemShut {NoStop}%
\bibitem [{\citenamefont {Affleck}\ and\ \citenamefont {Ludwig}(1993)}]{AffleckLudwig93}%
  \BibitemOpen
  \bibfield  {author} {\bibinfo {author} {\bibfnamefont {I.}~\bibnamefont {Affleck}}\ and\ \bibinfo {author} {\bibfnamefont {A.~W.~W.}\ \bibnamefont {Ludwig}},\ }\bibfield  {title} {\bibinfo {title} {Exact conformal-field-theory results on the multichannel kondo effect: Single-fermion green’s function, self-energy, and resistivity},\ }\href {https://doi.org/10.1103/PhysRevB.48.7297} {\bibfield  {journal} {\bibinfo  {journal} {Phys. Rev. B}\ }\textbf {\bibinfo {volume} {48}},\ \bibinfo {pages} {7297} (\bibinfo {year} {1993})}\BibitemShut {NoStop}%
\bibitem [{\citenamefont {Ishii}\ \emph {et~al.}(2013)\citenamefont {Ishii}, \citenamefont {Muneshige}, \citenamefont {Kamikawa}, \citenamefont {Fujita}, \citenamefont {Onimaru}, \citenamefont {Nagasawa}, \citenamefont {Takabatake}, \citenamefont {Suzuki}, \citenamefont {Ano}, \citenamefont {Akatsu}, \citenamefont {Nemoto},\ and\ \citenamefont {Goto}}]{PhysRevB.87.205106}%
  \BibitemOpen
  \bibfield  {author} {\bibinfo {author} {\bibfnamefont {I.}~\bibnamefont {Ishii}}, \bibinfo {author} {\bibfnamefont {H.}~\bibnamefont {Muneshige}}, \bibinfo {author} {\bibfnamefont {S.}~\bibnamefont {Kamikawa}}, \bibinfo {author} {\bibfnamefont {T.~K.}\ \bibnamefont {Fujita}}, \bibinfo {author} {\bibfnamefont {T.}~\bibnamefont {Onimaru}}, \bibinfo {author} {\bibfnamefont {N.}~\bibnamefont {Nagasawa}}, \bibinfo {author} {\bibfnamefont {T.}~\bibnamefont {Takabatake}}, \bibinfo {author} {\bibfnamefont {T.}~\bibnamefont {Suzuki}}, \bibinfo {author} {\bibfnamefont {G.}~\bibnamefont {Ano}}, \bibinfo {author} {\bibfnamefont {M.}~\bibnamefont {Akatsu}}, \bibinfo {author} {\bibfnamefont {Y.}~\bibnamefont {Nemoto}},\ and\ \bibinfo {author} {\bibfnamefont {T.}~\bibnamefont {Goto}},\ }\bibfield  {title} {\bibinfo {title} {Antiferroquadrupolar ordering and magnetic-field-induced phase transition in the cage compound prrh${}_{2}$zn${}_{20}$},\ }\href {https://doi.org/10.1103/PhysRevB.87.205106} {\bibfield  {journal}
  {\bibinfo  {journal} {Phys. Rev. B}\ }\textbf {\bibinfo {volume} {87}},\ \bibinfo {pages} {205106} (\bibinfo {year} {2013})}\BibitemShut {NoStop}%
\bibitem [{\citenamefont {Inui}\ and\ \citenamefont {Motome}(2020)}]{Motome_PhysRevB.102.155126}%
  \BibitemOpen
  \bibfield  {author} {\bibinfo {author} {\bibfnamefont {K.}~\bibnamefont {Inui}}\ and\ \bibinfo {author} {\bibfnamefont {Y.}~\bibnamefont {Motome}},\ }\bibfield  {title} {\bibinfo {title} {Channel-selective non-fermi liquid behavior in the two-channel kondo lattice model under a magnetic field},\ }\href {https://doi.org/10.1103/PhysRevB.102.155126} {\bibfield  {journal} {\bibinfo  {journal} {Phys. Rev. B}\ }\textbf {\bibinfo {volume} {102}},\ \bibinfo {pages} {155126} (\bibinfo {year} {2020})}\BibitemShut {NoStop}%
\bibitem [{\citenamefont {Hoshino}\ \emph {et~al.}(2013)\citenamefont {Hoshino}, \citenamefont {Otsuki},\ and\ \citenamefont {Kuramoto}}]{Hoshino_doi:10.7566/JPSJ.82.044707}%
  \BibitemOpen
  \bibfield  {author} {\bibinfo {author} {\bibfnamefont {S.}~\bibnamefont {Hoshino}}, \bibinfo {author} {\bibfnamefont {J.}~\bibnamefont {Otsuki}},\ and\ \bibinfo {author} {\bibfnamefont {Y.}~\bibnamefont {Kuramoto}},\ }\bibfield  {title} {\bibinfo {title} {Resolution of entropy \(\ln\sqrt{2}\) by ordering in two-channel kondo lattice},\ }\href {https://doi.org/10.7566/JPSJ.82.044707} {\bibfield  {journal} {\bibinfo  {journal} {Journal of the Physical Society of Japan}\ }\textbf {\bibinfo {volume} {82}},\ \bibinfo {pages} {044707} (\bibinfo {year} {2013})},\ \Eprint {https://arxiv.org/abs/https://doi.org/10.7566/JPSJ.82.044707} {https://doi.org/10.7566/JPSJ.82.044707} \BibitemShut {NoStop}%
\bibitem [{\citenamefont {Ge}\ and\ \citenamefont {Komijani}(2022)}]{Komijani_PhysRevLett.129.077202}%
  \BibitemOpen
  \bibfield  {author} {\bibinfo {author} {\bibfnamefont {Y.}~\bibnamefont {Ge}}\ and\ \bibinfo {author} {\bibfnamefont {Y.}~\bibnamefont {Komijani}},\ }\bibfield  {title} {\bibinfo {title} {Emergent spinon dispersion and symmetry breaking in two-channel kondo lattices},\ }\href {https://doi.org/10.1103/PhysRevLett.129.077202} {\bibfield  {journal} {\bibinfo  {journal} {Phys. Rev. Lett.}\ }\textbf {\bibinfo {volume} {129}},\ \bibinfo {pages} {077202} (\bibinfo {year} {2022})}\BibitemShut {NoStop}%
\bibitem [{\citenamefont {Tanaskovi\ifmmode~\acute{c}\else \'{c}\fi{}}\ \emph {et~al.}(2011)\citenamefont {Tanaskovi\ifmmode~\acute{c}\else \'{c}\fi{}}, \citenamefont {Haule}, \citenamefont {Kotliar},\ and\ \citenamefont {Dobrosavljevi\ifmmode~\acute{c}\else \'{c}\fi{}}}]{Darko_PhysRevB.84.115105}%
  \BibitemOpen
  \bibfield  {author} {\bibinfo {author} {\bibfnamefont {D.}~\bibnamefont {Tanaskovi\ifmmode~\acute{c}\else \'{c}\fi{}}}, \bibinfo {author} {\bibfnamefont {K.}~\bibnamefont {Haule}}, \bibinfo {author} {\bibfnamefont {G.}~\bibnamefont {Kotliar}},\ and\ \bibinfo {author} {\bibfnamefont {V.}~\bibnamefont {Dobrosavljevi\ifmmode~\acute{c}\else \'{c}\fi{}}},\ }\bibfield  {title} {\bibinfo {title} {Phase diagram, energy scales, and nonlocal correlations in the anderson lattice model},\ }\href {https://doi.org/10.1103/PhysRevB.84.115105} {\bibfield  {journal} {\bibinfo  {journal} {Phys. Rev. B}\ }\textbf {\bibinfo {volume} {84}},\ \bibinfo {pages} {115105} (\bibinfo {year} {2011})}\BibitemShut {NoStop}%
\bibitem [{\citenamefont {Haule}\ and\ \citenamefont {Kotliar}(2007)}]{Haule_PhysRevB.76.104509}%
  \BibitemOpen
  \bibfield  {author} {\bibinfo {author} {\bibfnamefont {K.}~\bibnamefont {Haule}}\ and\ \bibinfo {author} {\bibfnamefont {G.}~\bibnamefont {Kotliar}},\ }\bibfield  {title} {\bibinfo {title} {Strongly correlated superconductivity: A plaquette dynamical mean-field theory study},\ }\href {https://doi.org/10.1103/PhysRevB.76.104509} {\bibfield  {journal} {\bibinfo  {journal} {Phys. Rev. B}\ }\textbf {\bibinfo {volume} {76}},\ \bibinfo {pages} {104509} (\bibinfo {year} {2007})}\BibitemShut {NoStop}%
\bibitem [{\citenamefont {Sordi}\ \emph {et~al.}(2011)\citenamefont {Sordi}, \citenamefont {Haule},\ and\ \citenamefont {Tremblay}}]{Sordi-PhysRevB.84.075161}%
  \BibitemOpen
  \bibfield  {author} {\bibinfo {author} {\bibfnamefont {G.}~\bibnamefont {Sordi}}, \bibinfo {author} {\bibfnamefont {K.}~\bibnamefont {Haule}},\ and\ \bibinfo {author} {\bibfnamefont {A.-M.~S.}\ \bibnamefont {Tremblay}},\ }\bibfield  {title} {\bibinfo {title} {Mott physics and first-order transition between two metals in the normal-state phase diagram of the two-dimensional hubbard model},\ }\href {https://doi.org/10.1103/PhysRevB.84.075161} {\bibfield  {journal} {\bibinfo  {journal} {Phys. Rev. B}\ }\textbf {\bibinfo {volume} {84}},\ \bibinfo {pages} {075161} (\bibinfo {year} {2011})}\BibitemShut {NoStop}%
\bibitem [{\citenamefont {Gleis}\ \emph {et~al.}(2024)\citenamefont {Gleis}, \citenamefont {Lee}, \citenamefont {Kotliar},\ and\ \citenamefont {von Delft}}]{Gleis_PhysRevX.14.041036}%
  \BibitemOpen
  \bibfield  {author} {\bibinfo {author} {\bibfnamefont {A.}~\bibnamefont {Gleis}}, \bibinfo {author} {\bibfnamefont {S.-S.~B.}\ \bibnamefont {Lee}}, \bibinfo {author} {\bibfnamefont {G.}~\bibnamefont {Kotliar}},\ and\ \bibinfo {author} {\bibfnamefont {J.}~\bibnamefont {von Delft}},\ }\bibfield  {title} {\bibinfo {title} {Emergent properties of the periodic anderson model: A high-resolution, real-frequency study of heavy-fermion quantum criticality},\ }\href {https://doi.org/10.1103/PhysRevX.14.041036} {\bibfield  {journal} {\bibinfo  {journal} {Phys. Rev. X}\ }\textbf {\bibinfo {volume} {14}},\ \bibinfo {pages} {041036} (\bibinfo {year} {2024})}\BibitemShut {NoStop}%
\bibitem [{\citenamefont {Yao}\ and\ \citenamefont {Lee}(2011)}]{YL}%
  \BibitemOpen
  \bibfield  {author} {\bibinfo {author} {\bibfnamefont {H.}~\bibnamefont {Yao}}\ and\ \bibinfo {author} {\bibfnamefont {D.-H.}\ \bibnamefont {Lee}},\ }\bibfield  {title} {\bibinfo {title} {Fermionic magnons. non-abelian spinons, and the spin quantum hall effect from an exactly solvable spin-1/2 kitaev model with su(2) symmetry},\ }\href {https://doi.org/10.1103/PhysRevLett.107.087205} {\bibfield  {journal} {\bibinfo  {journal} {Phys. Rev. Lett.}\ }\textbf {\bibinfo {volume} {107}},\ \bibinfo {pages} {087205} (\bibinfo {year} {2011})}\BibitemShut {NoStop}%
\bibitem [{\citenamefont {Tsvelick}\ and\ \citenamefont {Wiegmann}(1985)}]{TsvW}%
  \BibitemOpen
  \bibfield  {author} {\bibinfo {author} {\bibfnamefont {A.~M.}\ \bibnamefont {Tsvelick}}\ and\ \bibinfo {author} {\bibfnamefont {P.}~\bibnamefont {Wiegmann}},\ }\bibfield  {title} {\bibinfo {title} {Exact solution of the multichannel kondo problem, scaling, and integrability},\ }\href {https://doi.org/0.1007/BF01017853} {\bibfield  {journal} {\bibinfo  {journal} {J. Stat. Phys.}\ }\textbf {\bibinfo {volume} {38}},\ \bibinfo {pages} {125} (\bibinfo {year} {1985})}\BibitemShut {NoStop}%
\bibitem [{\citenamefont {Hettler}\ \emph {et~al.}(1998)\citenamefont {Hettler}, \citenamefont {Tahvildar-Zadeh}, \citenamefont {Jarrell}, \citenamefont {Pruschke},\ and\ \citenamefont {Krishnamurthy}}]{DCA_PhysRevB.58.R7475}%
  \BibitemOpen
  \bibfield  {author} {\bibinfo {author} {\bibfnamefont {M.~H.}\ \bibnamefont {Hettler}}, \bibinfo {author} {\bibfnamefont {A.~N.}\ \bibnamefont {Tahvildar-Zadeh}}, \bibinfo {author} {\bibfnamefont {M.}~\bibnamefont {Jarrell}}, \bibinfo {author} {\bibfnamefont {T.}~\bibnamefont {Pruschke}},\ and\ \bibinfo {author} {\bibfnamefont {H.~R.}\ \bibnamefont {Krishnamurthy}},\ }\bibfield  {title} {\bibinfo {title} {Nonlocal dynamical correlations of strongly interacting electron systems},\ }\href {https://doi.org/10.1103/PhysRevB.58.R7475} {\bibfield  {journal} {\bibinfo  {journal} {Phys. Rev. B}\ }\textbf {\bibinfo {volume} {58}},\ \bibinfo {pages} {R7475} (\bibinfo {year} {1998})}\BibitemShut {NoStop}%
\end{thebibliography}%

 
\end{document}